\documentclass[11pt, oneside]{article}   	
\usepackage{geometry}                		
\usepackage{graphicx}				
\usepackage{amssymb}
\usepackage{amsmath}

\usepackage{authblk}

\usepackage{hyperref}

\title{Special Operations Forces: A Global Immune System?}
\author{Joseph Norman}
\author{\href{mailto:yaneer@necsi.edu}{Yaneer Bar-Yam}\vspace{-2ex}}
\affil{\href{http://www.necsi.edu}{New England Complex Systems Institute}, 210 Broadway Cambridge, MA 02139 \vspace{-2ex}}
\date{February 17th, 2016}

\begin{document}

\maketitle

\abstract{
The use of special operations forces (SOF) in war fighting and peace keeping efforts has increased dramatically in recent decades. A scientific understanding of the reason for this increase would provide guidance as to the contexts in which SOF can be used to their best effect, and when conventional forces are better suited. Ashby's law of requisite variety provides a scientific framework for understanding and analyzing a system's ability to survive and prosper in the face of environmental challenges. We have developed a generalization of this law to extend the analysis to systems that must respond to disturbances at multiple scales. This analysis identifies a necessary tradeoff between scale and complexity in a multiscale control system. As with Ashby's law, the framework applies to the characterization of successful biological and social systems in the context of complex environmental challenges. Here we apply this multiscale framework to provide a control theoretic understanding of the historical and increasing need for SOF, as well as conventional military forces. We propose that the essential role distinction is in the separation between high complexity fine scale challenges as opposed to large scale challenges. This leads to a correspondence between the role SOF can best serve and that of the immune system in complex organisms---namely, the ability to respond to fine-grained, high-complexity disruptors and preserve tissue health. Much like a multicellular organism, human civilization is composed of a set of distinct and heterogeneous \emph{social tissues}, each with its own distinct characteristics and functional relationships with other tissues. Responding to disruption and restoring health in a system with highly diverse local social conditions requires an ability to distinguish healthy tissue from disruptors and to neutralize disruptive forces with minimal collateral damage, an essentially complex task. Damage to social tissue, either through the growth of malignant forces or large-scale intervention by conventional forces, leads to cascading crises that spread beyond the initial location of disruption. To prevent such crises, the healthy functioning of social systems must be maintained by responding to disruptive forces while they remain small. SOF have the potential to mitigate against harm without disrupting normal social tissue behavior. Three conditions for SOF to fulfill such a role are identified: (1) distinctive capabilities of special operators that enable unmediated interaction with local cultures and peoples, (2)~persistent presence and embeddedness to foster cultural attunement and mutual trust, and (3) local autonomy and decision-making of SOF to achieve requisite variety for sensing and acting on fine-grained disturbances. We point out the inapplicability of traditional hierarchical control structures for high-complexity local tasks, which require a decentralized control architecture. This analysis suggests how SOF might be leveraged to support global stability and mitigate against cascading crises.
}

\newpage
\section{Executive Summary}
Special Operations Forces (SOF) provide war fighting capabilities that complement conventional forces. A conceptual framework is needed to clarify and differentiate the role of SOF within the larger military system to aid decision-makers in identifying when it is necessary and appropriate to utilize SOF and when conventional forces are better suited. 

Here, we propose a correspondence between the role SOF may serve and that of the immune system in complex organisms.

In organisms, the immune system is composed of many semi-autonomous components and is responsible for sensing and acting on fine-grained, high-complexity disturbances that may harm the growth and functioning of healthy tissue. It must differentiate between \emph{self} and \emph{other} by having an intimate knowledge of the character of local tissue, detecting agents and behavior that pose a threat. When functioning effectively the immune system eliminates harmful agents without disrupting the normal behavior of healthy tissue. 

Much like an organism, global civilization is composed of a collection of diverse \emph{social tissues}, each with its own distinct form, way of living, and functional role within larger communities. SOF are uniquely positioned to develop the knowledge and capabilities to distinguish healthy social tissue and detect and mitigate threatening forces. 

Conventional forces, by contrast, are well suited to external threats and their use in societal challenges may damage social fabric, leading to disrupted and vulnerable states. 

Three conditions must be met to enable SOF to eliminate threats while preserving the health of social tissue: 
\begin{enumerate}
	\item{Distinctive capabilities and advanced training of special operators -- Advanced cultural and language competencies and experience in making difficult decisions in the face of uncertainty enable unmediated interaction with local people. When necessary, they can strike with exacting force. }
	\item{Persistent presence and enduring engagements -- Repeated and habitual interaction with local communities at both the individual and institutional levels provides the opportunity to develop necessary cultural attunement.}
	\item{Local autonomy and decision-making -- Acting on nuanced information relevant to local conditions engenders the ability to stem local threats. Locally embedded SOF must have the freedom to behave semi-autonomously, making many decisions independently of SOF located elsewhere or central command structures. This requires avoiding the tendency to become bureaucratized.}
\end{enumerate}

The correspondence between the potential role of SOF and the immune system in organisms is formalized via multiscale control systems theory and a \emph{complexity profile} analysis. Ashby's \emph{law of requisite variety} sets the lower bound of complexity a system must possess to survive and prosper. SOF can provide essential complexity at fine scales, as does the immune system in biological systems. 

This correspondence suggests that SOF are uniquely equipped to serve as a global immune system, acting before threats rise to the level of crises, and preserving healthy and diverse social tissue functioning. Future policy decisions will determine the degree to which these unique SOF capabilities are developed and leveraged.

\section{Introduction}

Throughout history warfare has involved both large-scale conventional conflict, in which armed combatants seek to gain physical advantage over their adversaries, as well as less conventional operations which focus on high-value targets or seek to achieve a desired effect through indirect means. The latter have come to be known as \emph{special operations}, and they are typically carried out by small groups and individuals with distinctive skills, creativity, and often equipment. In 1987, the United States Special Operation Command (USSOCOM) was established to oversee the nation's special operations forces (SOF) for both independent and joint operations.

SOF have their origin in military practice, and only recently have attempts been made to articulate a theory of the role of SOF \cite{kiras2014theory, yarger201321st, spulak2007theory} as a subset of a larger \emph{theory of warfare}. These efforts highlight the need for a framework that provides guidance to decision-makers about when and how to utilize SOF to their greatest effect, and when other options are more appropriate. 

Here, we present a theory of SOF motivated by mathematical and physical necessity and grounded in complex systems science. We propose a correspondence between the functional role of SOF and that of the immune system in complex biological organisms, and a parallel correspondence between conventional forces and the neuromuscular system. According to this theory, SOF play a vital role in sensing and acting in fine-grained, high-complexity environments, complementing conventional forces that sense and act at larger scales.

The theory brings military theory into contact with a body of scientific knowledge and inquiry about the behavior of complex systems and, crucially, the conditions under which they are able to survive and prosper. 

The remainder of the article is divided into four sections. First, relevant concepts in the theory of multiscale control systems are reviewed and summarized. Second, these concepts are applied to clarify the functional complementarity of the immune and neuromuscular systems in complex organisms. Third, the functional role of SOF is couched in this theory and brought into correspondence with that of the immune system. Finally, strategy, policy implications, and implementation challenges are discussed. 

\section{Multiscale Control Systems}

\subsection{The law of requisite variety and its limitations}

\begin{figure}[h]
\centering
\includegraphics[width=.75\linewidth]{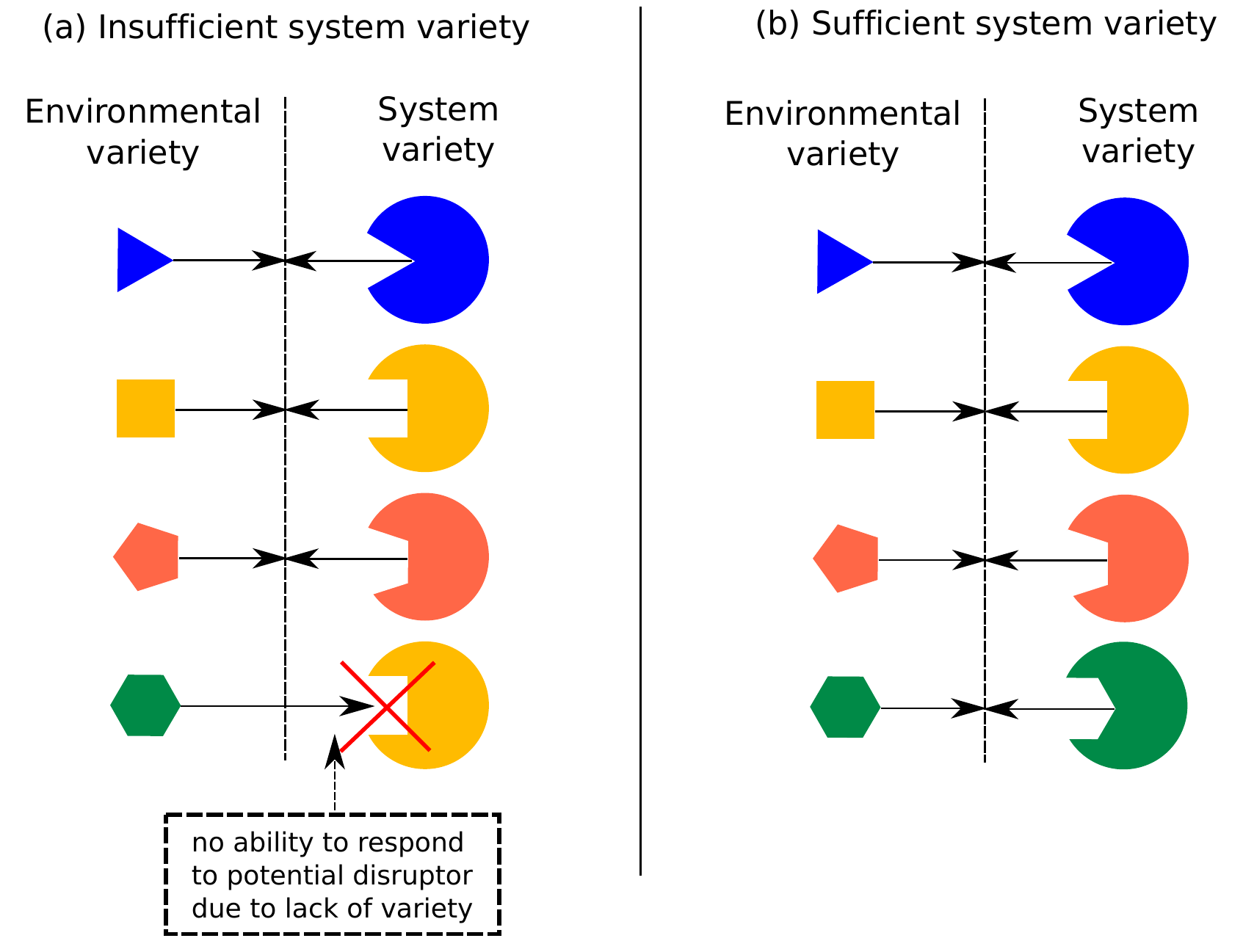}
\caption{\scriptsize{Requisite Variety. Panel (a) shows a system (right) being exposed to environmental disturbances (left). The variety of the environment is greater than the variety of the system, as there are 4 unique disturbances but only 3 unique responses. The system lacks the ability to respond to the green hexagon, which will disrupt it. Panel (b) shows a similar case, but where the system variety matches the variety of potential disturbances---the system has the requisite variety to respond to all potential disturbances. Systems with variety greater than that of their environment also possess requisite variety. }}
\label{Fig:RV}
\end{figure}

In 1956, W. Ross Ashby formalized in the study of control systems what is known as \emph{the law of requisite variety} \cite{ashby1956introduction}. In short, the law of requisite variety sets the minimum number of behaviors, or `variety', a system must have to survive and prosper in a given environment. As the number of distinct situations a system encounters increases, the variety of its behavioral repertoire must also increase in order to achieve desired outcomes---or as Ashby put it: ``variety destroys variety''. This concept is illustrated in Figure \ref{Fig:RV}. If a system has little variety or is overly-constrained while being exposed to a large variety of stressors (i.e. a complex environment), it will sooner or later fail to achieve desired outcomes. In this article, we will use the terms \emph{variety} and \emph{complexity} interchangeably. Thus an environment with \emph{high-complexity} is one with a large variety. 

The theory of control systems traditionally deals with systems at a well-defined \emph{scale} of relevant behavior, and abstracts away details that are presumed not to be of concern due to the nature of the system or the method of control. For example, if one wanted to construct a robot that could catch a baseball, one need not be concerned with the atomic vibrations ongoing within the baseball, but rather its relevant macroscopic properties like mass, location, and trajectory, and corresponding control variables like joint angles and positions.

In contrast, living systems are exposed to environments with stressors and complexity on multiple relevant scales that must be effectively managed to achieve self-regulation and good overall system health. For example, as organisms we are exposed not only to traffic as we cross the street, but also to microscopic organisms that may find our bodies to be suitable homes within which to replicate themselves to our detriment. These two sources of stress exist at scales separated by several orders of magnitude, and our bodies therefore have different strategies in controlling for their potentially harmful effects. 

Thus, the law of requisite variety per se is not enough to account for how, say, an organism achieves self-regulation in a complex environment with multiple scales of impinging forces and stressors. Both \emph{variety} and \emph{scale} must be considered for good control in complex multiscale environments \cite{bar2004multiscale, allen2014information}. 

\subsection{Scale / complexity tradeoff}

There is an inherent tradeoff between the scale and complexity of behavior in any system. In order for large-scale behaviors to occur, a large number of components must work coherently or in coordination. Consider, for example, the muscle tension that ultimately gives rise to the movement of a limb. If only one or a small number of muscle fibers become engaged, the scale of the force will be small, and the limb will express essentially no behavior. However if many muscle fibers become engaged at once, a larger scale force is produced, and the limb will change its position---a large-scale behavior is induced through the coherent activity of many parts. 

The flip-side of achieving large-scale effects through coherent behavior of many components is that those components are not free to behave independently, but are constrained by the role they play in the large-scale behavior. This decreases the variety, or complexity, that can be expressed by the system at small scales.

We can quantify a system's variety at a given scale. In a system with $N$ components that behave independently, the number of states the system can achieve is the product of the number of states each component can take. For instance if each component can take on $2$ states (a \emph{binary} system), the number of total possible states is $2^{N}$. More generally, if $n_i$ denotes the number of states component $i$ can take on, the total number of states of the system is $\prod_i n_i $. The components must act in concert in order to achieve large-scale effect. This necessarily reduces their degree of independence, and the number of states of the system, or variety, is less than $\prod_i n_i$. In other words, due to constraints that prevent each component from behaving independently, the actual variety is less than its maximum would be without constraint. Constraints are indicative of underlying structures that enable large-scale behaviors and variety. 

We can summarize the essential tradeoff as follows: \emph{demand for variety at large-scales necessitates the reduction in variety at smaller scales.}

\begin{figure}[b!]
\centering
\includegraphics[width=.85\linewidth]{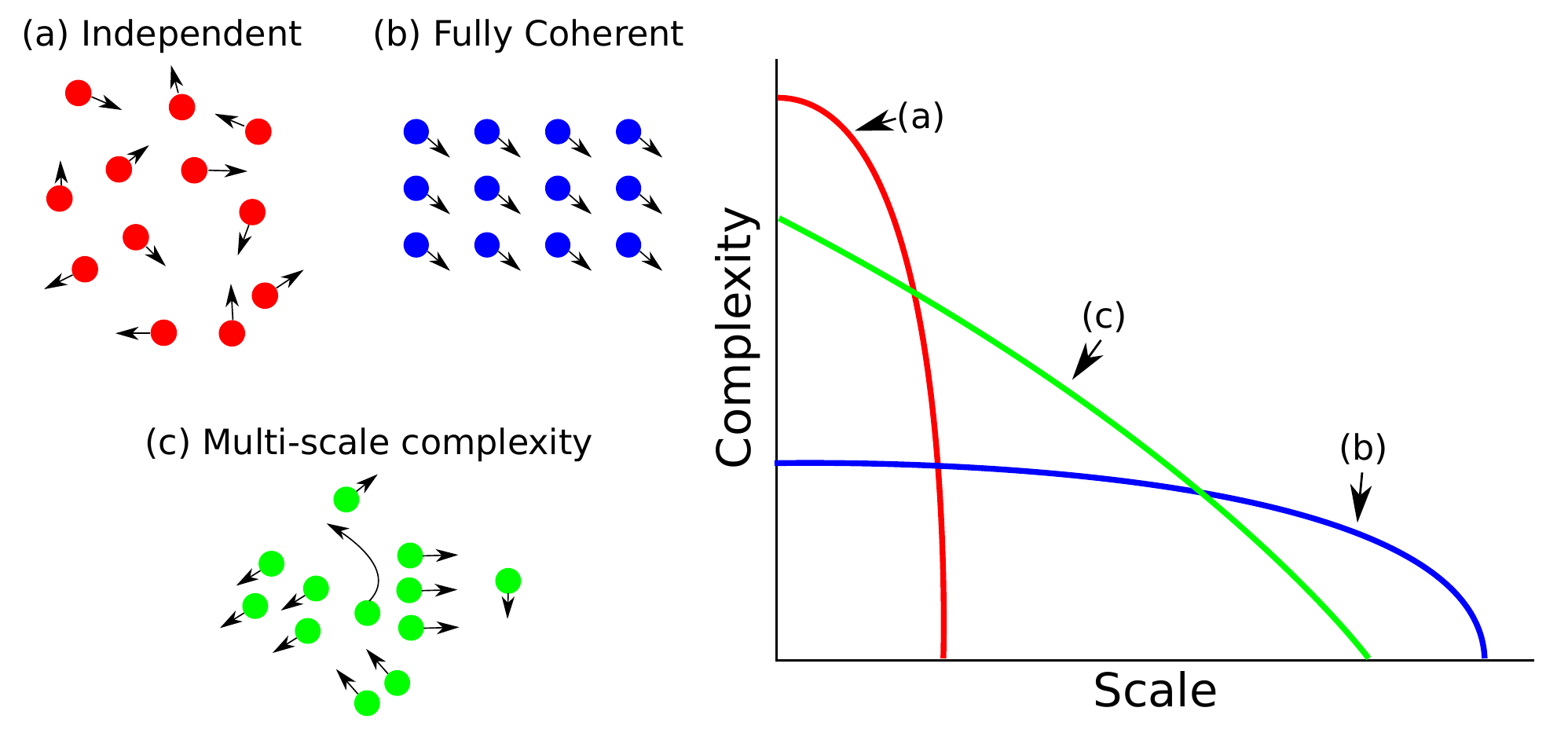}
\caption{\scriptsize{Complexity Profile. The complexity profile maps system complexity, or variety, as a function of scale. Three example cases are shown. In the case of system components behaving with complete independence (a), there is high-complexity at the finest scales, but variety quickly drops off to zero, and no behavior is observed at larger scales. A system in which the components are fully coherent (b) has substantially reduced variety at small scales. Intuitively, this is because if we know the behavior of one component, we know them all (i.e. they are constrained). However, this enables large-scale behavior, as the components behave in concert.  Systems with multiscale complexity (c) have both fine-scale complexity (though less than case (a)) and can produce large scale behaviors (though not as large-scale as case (b)). This is achieved by some components behaving in a coherent and coordinated fashion, while others are free to behave semi-autonomously.}}
\label{Fig:CP}
\end{figure}

For a given system, this tradeoff can be captured and summarized via the \emph{complexity profile} (Figure \ref{Fig:CP}) which represents variety, or complexity, as a function of scale \cite{allen2014information, bar2013computationally, bar2001multiscale, bar2004making, bar1997dynamics}. The `shape' that the complexity profile of a system takes on reflects its structure and behavior and identifies the scales over which they are present. When the smallest components of a system behave essentially independently, there is a maximal amount of variety at the fine-scale. However, as we move to larger scales, the independent behavior of all these parts `average out' and we observe no large-scale behavior. When all of the smallest components move together coherently across the entire system, like the atoms in a baseball when thrown, we find behavior at larger scales, with variety varying minimally across scales, and variety at fine-scales being reduced dramatically compared to the case of component independence. If you know the flight path of one atom in the baseball, you know them all. For objects like complex organisms, we observe a mixture of these two modes. The variety at the smallest scales remains quite high (though, less than in the case of complete independence), while many of the components are coordinated into larger structures that reduce their independence, but achieve larger scale behaviors. 

The different multiscale behaviors also require different control structures to enable actions to be performed in response to environmental challenges or conditions \cite{bar2004multiscale}. For large scale behavior, hierarchical control is appropriate. This is because, on the sensing side, large, coherent external events are detected, and irrelevant details are filtered out for high-level decision making. On the action side, unified decisions can be projected to a large number of agents who behave in concert to achieve a large-scale effect. In contrast, for behavior that responds to fine-scaled, high-complexity challenges that don't require large-scale response, hierarchical control is inappropriate and insufficient. The details lost as information ascends the hierarchy are precisely the ones relevant to small scale decisions. Moreover, the projection of operational directives from high-levels is necessarily insensitive to these low-level details, constraining agents' behavior and preventing them from responding and adapting to the local context. Instead, distributed networks of agents with minimal hierarchical constraints leave small groups and individuals able to make decisions semi-autonomously, retaining sensitivity to local information and enabling adaptation and response to these locally-relevant variables. 

\section{The immune and neuromuscular systems}

A familiar example of a multiscale control system with clear differentiation of scale and function is the physiological system that combines the neuromuscular and immune systems in multicellular organisms, such as ourselves. The key differences between the form and function of these two systems lend insight into the nature and role of SOF, and that of conventional forces.

The neuromuscular and immune systems operate concurrently in order to achieve overall system health by responding to disturbances at different scales. The neuromuscular system detects large-scale events and structures in the external environment---dodgeable cars, walkable paths, climbable trees, fall-offable cliffs---and generates coordinated behavior of the gross physical structure of the body in order to leverage opportunities or mitigate harm. These faculties operate in the `Newtonian' macroscopic environment of everyday life to avoid physical damage and provide the resources necessary for physiological function. 

The immune system serves a different, but equally important, function. It is distributed and embedded throughout the body and its tissues. It contains a variety of cell types that behave with a large degree of independence---behaviors are not constrained to achieve large-scale coherence as in the neuromuscular system. One of its essential roles is to differentiate \emph{self} and \emph{other} at the cellular and sub-cellular scales in order to promote the flourishing of healthy tissue and eliminate or neutralize threats when detected. The cells of the immune system sense and act locally, without direct instruction from centralized command structures, though `training' and other functions are centralized in lymph nodes and bone marrow. 

The distinction between self and other is not genetic, but rather associated with healthy functioning. Consider a cancerous cell and a bacterium that aids in healthy digestion. The former would be appropriately identified as \emph{other}, despite sharing its genome with the host, and the latter \emph{self}, because of its functional harmony with the host. 

The body is organized into a collection of heterogeneous tissues and organs which serve various functions that complement one another, forming a self-consistent whole. A well-functioning immune system promotes a healthy system by minimizing the potential for disruption of local tissue. This is a critical point: the integrity of the functional tissue is preserved via the action of a healthy immune system, and a disruption to \emph{any} of the tissues in the body disrupts their role and can lead to cascading effects throughout the whole system, including those cells not directly affected by the disruptor, and perhaps organism death. 

Notably, the immune system does not \emph{direct} the tissue or instruct its behavior explicitly, but rather creates the conditions in which it can express its distinct form without harming itself or other tissue. 

The essential differences between the neuromuscular and immune systems are the scales over which they operate, and the degree of independence of components that determine the scale and variety. The neuromuscular system operates with a great degree of coordination among its parts, limiting variety at fine-scales and producing it at large-scales. The components of the immune system, by contrast, behave more independently, making decisions locally and maintaining variety at the fine-scales, enabling sensing and acting that preserves good tissue functioning and avoids disruption. 

The immune system cannot catch a baseball, and the neuromuscular system cannot eliminate a bacterium. The \emph{only} way the neuromuscular system could effectively combat a micro-disruptor would be through the destruction of functional tissue, an action with irreversible and often system-wide consequences.

\section{Multiscale military theory and the functional role of SOF}

Much like an organism, our global civilization is composed of a set of distinct \emph{social tissues}, each with unique character, mode of internal operation, and interfaces with other tissues. Healthy, well-functioning social tissues have internal behaviors that sustain the individuals composing them, such as agriculture, goods production, trading and markets, health services, social gatherings and celebrations, as well as fruitful external interactions with other social systems such as the buying and selling of commodities, products, and services. 

When healthy and functional social tissue is disrupted, opportunities are created for malignant forces to gain footholds and grow. This dynamic can be seen, for instance, in the unintended consequences of the invasion of Iraq, which created the opportunity for terrorist networks and other harmful actors to increase their power and influence as normal life was disrupted and power vacuums were created. Moreover, the harm and risks generated by the growth of malignant forces are not confined to the local area where they first manifest. 

Because of our global interconnectedness and interdependence, effects cascade causing disruption in other tissues, leading to a domino effect with no straightforward mechanism to halt the expanding impacts \cite{bar2002complexity, lagi2015accurate}. The recent and ongoing migrant crisis in Europe and beyond provides an example of one form such cascading effects can take. 

Global interdependency means any large-scale military intervention, by virtue of disrupting the normal functioning of society, will generate both local and non-local unintended consequences even when desired effects are achieved. This is not to suggest that large-scale action is never necessary, but the potential for generating new crises must be weighed carefully whenever it is considered as an option. In many cases, action that does not disrupt local, healthy social behavior is possible, but it requires the right action and agent. 

The parallels of the effects of tissue disruption in organisms and in sociocultural systems highlights the need for a `sociocultural immune system'---a fine-grained system for sensing and acting on environmental disturbances at scales smaller than conventional forces are able. In this regard, conventional forces can be likened to the large-scale neuromuscular system in organisms. Acting instead at a small-scale presents the possibility of maintaining healthy social tissue and allowing it to flourish. Just as for the immune system, this is not a matter of differentiating `native' and `foreign', but understanding whether an agent is disruptive to overall health. 

SOF are uniquely positioned to fulfill this role, possessing the requisite personnel, skills, and training. For this to be realized, policies that impact SOF must be such that they enable their unique capabilities in meeting the high-complexity demand of local cultural systems. We identify three conditions that must be satisfied in order for SOF to serve such a role: special operators with advanced training and distinctive capabilities, persistent presence and enduring engagements, and local autonomy and decision-making. We discuss each in turn. 

\subsection{Distinctive capabilities}

Much like the cells in the immune system have special forms and functions to fulfill their roles, the distinctive capabilities of special operators enable them to operate in highly complex sociocultural environments. Advanced language and cultural training allows unmediated interaction with local peoples. Special operators' experience in making decisions in the face of uncertainty allow them to operate in ill-defined `gray zone' conditions. 

The need to produce special operators with distinctive capabilities highlights the role of SOF's high-selectiveness, and emphasizes the necessity of advanced training in language and culture in addition to combat. These values are articulated in the SOF truths ``Humans are more important than hardware'' and ``SOF cannot be mass produced" \cite{SofTruths}. Preparing special operators to interact directly and make difficult decisions in complex psychosocial, sociocultural, and kinetic environments must be a priority of SOF and their enabling agencies. 

\subsection{Persistent presence}

The immune system is embedded throughout the tissues of the body to develop and maintain sensitivity to the character of local tissue and respond rapidly to disruptors \cite{matzinger2002danger}. Similarly, persistent presence of SOF allows for nuanced relationships to unfold over time, and for cultural attunement to be developed at both the individual and institutional levels. SOF embeddedness engenders an understanding of normal conditions and a sensitivity to changes in those conditions and whether they pose a threat. Moreover, presence is necessary for applying rapid and effective action to achieve desired effects with minimal disruption. Just as SOF must recognize `self' in multiple contexts, local cultures must not react to SOF as a foreign entity, i.e. mutual trust must be present, developed through shared history.

Policies should enhance continuity of interaction between SOF and a given sociocultural system even, or especially, when there is no immediate or visible threat. The only way to prevent the growth of malignancies is to be present and active \emph{before} they grow. This is reflected in the SOF truth ``competent SOF cannot be created after emergencies occur" \cite{SofTruths} and Admiral William McRaven's oft-cited comment that one ``can't surge trust'' \cite{mcraven}.

\subsection{Local autonomy and decision-making}

As the cells of the immune system sense, decide, and act locally in a decentralized manner, being fine-tuned to the character of their local tissues, so too must SOF have the ability to sense, decide, and act locally using their nuanced understanding and experience. 

The semi-autonomy of SOF is necessary for requisite variety to be achieved in interfacing with high-complexity, fine-grained environments and disruptors.

In human systems, these disruptors manifest at the psychosocial and sociocultural scales. It is possible to take effective action at these scales to eliminate harmful agents without disrupting healthy social tissue functioning. This becomes impossible as the scale of a malignancy grows larger: social tissue will inevitably be damaged by both the malignancy itself and any large-scale force applied in response. 

When the decision-making agent is both far removed from and insensitive to the  the local context, as well as receiving multiple information streams about which decisions must be made, the sensitivity, nuance, and understanding of local SOF is lost. Consequently, the ability to stem malignant forces while they remain small in scale is diminished, and the likelihood of disrupting a social system either accidentally or out of necessity as the scale of harmful actors grows larger increases.

To enable SOF to act without disrupting social tissue, the institutions overseeing SOF must not over-constrain their behavior. As policy- and decision-makers look increasingly to SOF to overcome complex challenges, it is critical that they do not become overly-bureaucratized. 

Imperatives that are communicated to SOF must be guided by their role as a protector of local tissue function. Protections from the potential for harm to local tissues, i.e. by civilian collateral damage from operations, must be instituted in a way that retains local autonomy. Detailed instructions on how to carry out missions will prevent them from behaving as necessary for success in high-complexity environments. In technical terms, placing too many constraints on their behavior will reduce their variety below the (requisite) threshold for sensing and acting on fine-grained disruptors. The consequences of this are twofold: (1) SOF will lack the ability to sense and eliminate threats while they remain small, and (2) disruption and destruction of healthy social tissue becomes inevitable as malignant forces grow and large-scale intervention becomes the only means of engagement. 

\section{Challenges and Implementation}

While SOF are uniquely positioned to fulfill an immune-system-like function, there remain significant challenges to successful implementation. Here we summarize some of these challenges.

Developing SOF who are both culturally and linguistically competent, as well as able to execute reconnaissance and surveillance and direct action missions demands significant investment in training and preparation. Moreover, for any individual operator there is a tradeoff in developing proficiency in any given domain. However, SOF must be able to perform the entire range of activities, from sensing nuanced changes in social conditions to taking actions to eliminate harmful disruptors, to preserving social tissue health.

Fundamental limitations on individual capabilities lead to a need for diversification of roles of SOF. This is manifest already in different types of SOF, as it is in the immune system which uses cells of various types, each of which serves particular roles that complement one another. The relative levels of activity for the different cell-types vary depending on circumstance; some cells primarily sense tissue conditions and detect disruptors, while others act to confine and eliminate harmful agents once they are identified. During an infection rapid clonal reproduction (replication) of effective types occurs. Similarly, SOF may embrace and develop specialization of expertise, and should be flexible enough to adapt force size and composition in response to changing circumstances.

Maintaining the mental health of special operators must be a priority, and appropriate support systems should be put in place for this. The high complexity of tasks translates into the psychological symptoms of stress, depression and burnout, common in a high complexity society more generally but surely for SOF. Moreover, adapting to diverse local contexts creates challenges when switching to home and family environments, a potential component of post traumatic stress disorder (PTSD). This is a challenge for both the SOF and their families.

Giving special operators a significant degree of autonomy presents challenges and risks that are distinct from those of conventional command-and-control systems. Care must be taken to ensure social tissue is not damaged by unintentional friendly fire, collateral damage or intentional `rogue operators'. The potential for disfunction is not unlike auto-immune disorders in complex organisms and the immune system has developed mechanisms for prevention, though no mechanism is failure proof. Local feedback systems including multiple specialized roles rather than centralized control ones must be in place that put checks on the actions of operators.The structure of these feedback mechanisms must be the subject of intensive study. 

Rapid growth in recent years has led to institutionalization of SOF using concepts that may be incorrectly adopted from command control military traditions. Bureaucratization runs counter to the ability of SOF for performing the functions we have identified. Rather then enabling SOF function as it grows, institutionalization may result in undermining the effectiveness of SOF as it becomes more like conventional forces. Alternative structures must be developed. They may be inferred from fundamental complex systems analyses, including correspondence with immune system functions or well designed experimentation.

Institutional structures and relationships between SOCOM and other enabling agencies, including those within DoD, and other departments of the executive branch such as the State Department, need to be carefully considered. For example, how the agenda of an ambassador of a given region and local SOF should interrelate is an open question. If command and control structures do not appropriately interface with SOF, their unique capabilities will not be utilized effectively. This includes knowing when and, crucially, when \emph{not} to utilize SOF to achieve a desired effect. This article is intended to contribute to this clarification.
\section{SOF in the 21st century}

There is no doubt that as a global civilization we will continue to face fine-grained, high-complexity disruptors that have the potential to grow into larger-scale malignancies. The only way to combat this is to promote and enable the flourishing of healthy social tissues. Multiscale control systems theory makes clear the need for an immune-like system embedded within human social systems. It must be sensitive to and embedded within high-complexity psychosocial and sociocultural environments to make decisions locally based on understanding of a given social system, its nuances, and distinctive qualities. 

Like the various tissues arranged into functional organs throughout the body, cultures and social systems do not all look, behave, or function alike. Part of a global strategy for the 21st century must be the recognition that cultures can not simply be `exported' or `projected' onto others without pushback, and that behavioral diversity at the collective scale is a natural and healthy part of our human civilization. SOF possess the unique organizational capabilities to be sensitive to the healthy behavior of these diverse `social tissues', while providing the direct and indirect action capabilities to neutralize malignant forces when identified. 

Moving forward, a major part of the SOF repertoire must include relationship building. Interpersonal relationships with local individuals form the basis of understanding necessary to discern between harmful and beneficial (or neutral) forces to social health. The ability to perceive and understand local tensions, grievances, typical and atypical interactions, customs, and other nuanced features can serve to generate solutions before the normal functioning of healthy social tissue is threatened. The highly-complex and fine-grained nature of this endeavor makes it an unsuitable role for conventional forces -- they can not sense nor act on such a fine scale. A focus on direct action is important when specific disruptors have been identified, and not otherwise. 

SOF is uniquely positioned to serve as a global immune system, keeping the diverse set of social tissues healthy, and reserving large-scale intervention for when it is necessary. 

\vspace{5mm}
\noindent We thank Charles Flournoy and Philip Kapusta for helpful comments and discussion.

\bibliography{SoImmune}
\bibliographystyle{necsi}

\end{document}